\documentstyle[aps,multicol,prl,epsf]{revtex}
\begin{document}
\draft

\title{Measuring the transmission of a quantum dot using Aharonov-Bohm Interferometers}

\author{%
Amnon Aharony$^a$, Ora Entin-Wohlman$^a$,
and Yoseph Imry$^b$}

\address{%
$^a$ School of Physics and Astronomy, Raymond and Beverly Sackler
faculty of Exact Sciences, \\Tel Aviv University, Tel Aviv 69978,
Israel\\
$^b$ Department of Condensed Matter Physics, Weizmann Institute of
Science, Rehovot 76100, Israel.}

\date{\today}
\maketitle

\begin{abstract}
 The conductance ${\bf G}$ through a {\bf closed}
Aharonov-Bohm mesoscopic solid-state interferometer (which
conserves the electron current), with a quantum dot (QD) on one of
the paths, depends only on $\cos\phi$, where $\Phi=\hbar c \phi/e$
is the magnetic flux through the ring. The absence of a phase
shift in the $\phi-$dependence led to the conclusion that closed
interferometers do not yield the phase of the ``intrinsic"
transmission amplitude $t_D=|t_D|e^{i\alpha}$ through the QD, and
led to studies of {\bf open} interferometers. Here we show that
(a) for single channel leads, $\alpha$ can be deduced from
$|t_D|$, with no need for interferometry; (b) the explicit
dependence of ${\bf G}(\phi)$ on $\cos\phi$ (in the closed case)
allows a determination of {\it both} $|t_D|$ {\it and} $\alpha$;
(c) in the open case, results depend on the details of the
opening, but optimization of these details can yield the two-slit
conditions which relate the measured phase shift to $\alpha$.

\end{abstract}

\begin{multicols}{2}

\section{Introduction}

Recent advances in nanoscience raised much interest in quantum
dots (QDs), which represent artificial atoms with experimentally
controllable properties\cite{review,book}. The quantum nature of
the QD is reflected by resonant tunneling through it, as measured
when the QD is connected via metallic leads to electron
reservoirs. The measured conductance ${\bf G}$ shows peaks
whenever the Fermi energy of the electrons crosses a resonance on
the QD. Experimentally, the energies of these resonances are
varied by controlling the plunger gate voltage on the QD, $V$.
Quantum mechanically, the information on the tunneling is
contained in the complex transmission amplitude,
$t_{D}=\sqrt{{\cal T}_{D}}e^{i\alpha}$. It is thus of great
interest to measure both the magnitude ${\cal T}_{D}$ and the
phase $\alpha$, and study their dependence on $V$. Although the
former can be deduced from measuring ${\bf G}$, via the Landauer
formula \cite{landauer}, ${\bf G}=\frac{2e^2}{h}{\cal T}$,
experimental information on the latter has only become accessible
since 1995\cite{yacoby,schuster}, using the Aharonov-Bohm (AB)
interferometer\cite{azbel}.

In the AB interferometer, an incoming electronic waveguide is
split into two branches, which join again into the outgoing
waveguide. Aharonov and Bohm\cite{AB} predicted that a magnetic
flux $\Phi$  through the ring would add a difference
$\phi=e\Phi/\hbar c$ between the phases of the wave functions in
the two branches of the ring, yielding a periodic dependence of
the overall transmission ${\cal T}$ on $\phi$. Placing a QD on one
of the branches, as in Fig. 1a, and using the other path as a
``reference path", with a transmission amplitude $t_B$, one
expects ${\cal T}$ also to depend on the {\it ``intrinsic"
amplitude} $t_{D}$. In the two-slit limit, one has  ${\cal
T}=|t_De^{i\phi}+t_B|^2=A+B\cos(\phi+\beta)$, with
$\beta=\alpha+\kappa$, where the reference phase $\kappa$ is
independent of the QD parameters, and thus set at zero. However,
for the ``closed" two-terminal geometry, as shown in Fig. 1a and
used by Yacoby {\it et al.}\cite{yacoby}, the two-slit expectation
that $\beta=\alpha$ was clearly wrong: Unitarity (conservation of
current) and time reversal symmetry imply the Onsager relations
\cite{onsager,but}, which state that ${\bf G}(\phi)={\bf
G}(-\phi)$, and therefore $\beta$ {\it must} be equal to zero or
$\pi$. Indeed, a fit of the the experimental data\cite{yacoby} to
the above two-slit formula, with $B>0$, gives a phase shift
$\beta$ which ``jumps" from 0 to $\pi$ near each resonance of the
QD, and then exhibits an a priori unexpected abrupt ``phase lapse"
back to 0, between every pair of resonances.

\begin{figure}[h]
\begin{center}
\leftline{\epsffile{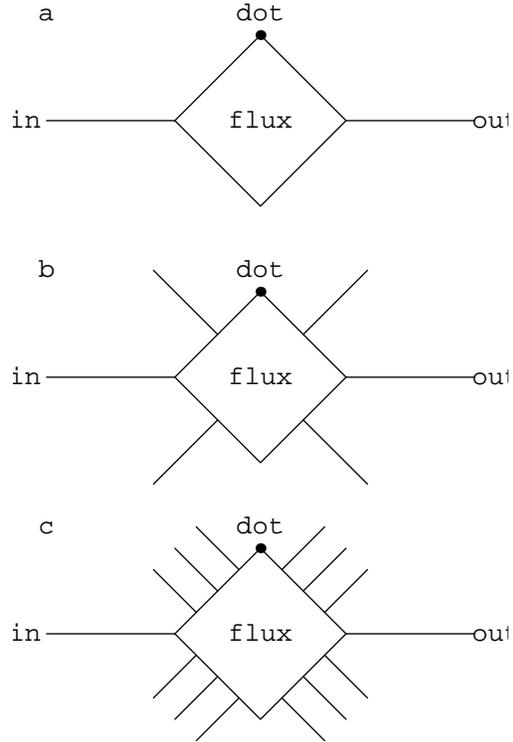}} \caption{Model for the AB
interferometer: (a) Closed two-terminal case, (b) schematic
picture of the six-terminal open interferometer, (c) model for the
open interferometer.} \label{AB}
\end{center}
\end{figure}

Aiming to measure a non-trivial AB phase shift $\beta$ then led to
experiments with six-terminal ``open" interferometers (Fig.
1b)\cite{schuster,ji}, where the additional four ``leaky" channels
lead to absorbing reservoirs. These interferometers break the
Onsager symmetry, and yield a non-trivial phase shift $\beta$
which increases {\it gradually} from zero to $\pi$ through each
resonance, and then jumps back to zero between resonances. Much of
the early literature {\bf assumed} that the measured $\beta$ is
equal to the ``intrinsic" $\alpha$. Recently \cite{prl} we showed
that this assumption is not necessarily always valid: the detailed
$V$-dependence of $\beta$ {\bf depends on the strength} of the
coupling to the additional terminals. Thus, we posed the challenge
of finding clear criteria as to when the measured $\beta$ equals
$\alpha$.

In the present paper we review three aspects of this problem. In
Sec. 2 we show that in some cases one has ${\cal
T}_D=\gamma_D^2\sin^2\alpha$, where $\gamma_D$ is a constant
measuring the asymmetry of the QD. In these cases, $\alpha$ is
determined by ${\cal T}_D$, and there is no need to build special
interferometers to measure $\alpha$. We also show that in many
cases the detailed flux-dependence of the measured conductance of
the {\bf closed} interferometer contains information which allows
the deduction of both ${\cal T}_D$ and $\alpha$, eliminating the
need to open the interferometer. In Sec. 3 we then review recent
work \cite{bih} in which we showed that an opening like that shown
in Fig. 1c, with ``forks" of ``lossy" channels on each segment of
the AB ring, can be tuned so that one reproduces the two-slit
conditions. In those cases, one indeed has $\beta=\alpha$.
However, this tuning requires some optimization of the parameters,
and cannot be guaranteed for an arbitrary open interferometer.

\section{Closed AB interferometer}
\subsection{Model with one QD resonance}

Our model is shown in Fig. 2: We start with an isolated quantum
dot, with a single localized level (of single particle energy
$\epsilon_{D}$, representing the gate voltage $V$) and on-site
Hubbard interaction $U$, and with a one-dimensional tight-binding
chain, with sites at integer coordinates, with zero on-site
energies and with nearest-neighbor (nn) hopping matrix elements
equal to $-J$. We next connect the dot to the sites $\pm 1$ on the
chain, via matrix elements $-J_{\ell}$ to the left, and $-J_{r}$
to the right, and also modify the lower branch of the resulting
triangle: site 0 becomes the ``reference" site, with on-site
energy $\epsilon_0$, with no interactions and with  nn hopping
energies $-I$ (replacing the original $-J$). Experimentally, the
reference site can represent a simple point contact, tunnel
junction, etc. The triangle so formed contains an Aharonov-Bohm
phase, $\phi =\phi_{\ell}+\phi_{r}$, where (for simplicity), the
phase $\phi_{\ell}$ ($\phi_{r}$) is attached to the bond with the
hopping matrix element $J_{\ell}$ ($J_{r}$) (we choose
$J,~I,~J_\ell$ and $J_r$ to be real). Hence, the Hamiltonian of
the system reads
\begin{eqnarray}
{\cal H}&={\cal
H}_{D}+\sum_{k\sigma}\epsilon_{k}c^{\dagger}_{k\sigma}c_{k\sigma}
+\epsilon_{0}\sum_{\sigma}c^{\dagger}_{0\sigma}c_{0\sigma}\nonumber\\
&+(I/J-1)\sum_{k\sigma}\epsilon_{k}\Bigl
(c^{\dagger}_{k\sigma}c_{0\sigma}+c^{\dagger}_{0\sigma}c_{k\sigma}\Bigr
)/\sqrt{N}\nonumber\\
&+\sum_{k\sigma}\Bigl ({\cal
V}_{k}d^{\dagger}_{\sigma}c_{k\sigma}+{\cal
V}^{\ast}_{k}c^{\dagger}_{k\sigma}d_{\sigma}\Bigr ), \label{HH}
\end{eqnarray}
where the operator $c_{k\sigma}^\dagger$ creates single particle
eigenstates (with spin $\sigma$) on the unperturbed chain (with
$I=J$, $J_\ell=J_r=0$), with eigenenergy $\epsilon_{k}=-2J\cos k$,
while
$c_{0\sigma}=\sum_{k}c_{k\sigma}/\sqrt{N}$,
 and
${\cal V}_{k}=-(J_{\ell}e^{i\phi_{\ell}-ik}+J_{r}e^{-i\phi_{r}+ik}
)/\sqrt{N}$.
The operators on the dot are denoted by $d_{\sigma}$ and $d^{
\dagger}_{\sigma}$, and they anti-commute with the band operators
$c_{k\sigma},c^{\dagger}_{k\sigma}$. The dot Hamiltonian is
\begin{eqnarray}
{\cal
H}_{D}=\epsilon_{D}\sum_{\sigma}d^{\dagger}_{\sigma}d_{\sigma}+
\frac{1}{2}U\sum_{\sigma}n_{d\sigma}n_{d\overline{\sigma}},
\end{eqnarray}
with
$n_{d\sigma}=d^{\dagger}_{\sigma}d_{\sigma}$,
and $\overline{\sigma}$ denotes the spin opposite to $\sigma$. The
Hamiltonian (\ref{HH}) is a simple generalization of that used by
Ng and Lee\cite{ng}, to which we added the reference path.

\begin{figure}[h]
\begin{center}
\leftline{\epsfclipon\epsfxsize=3in\epsfysize=2in\epsffile{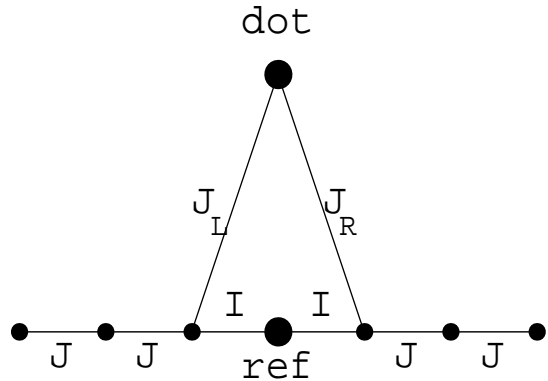}}
\caption{Model for the closed interferometer.}\label{fig2}
\end{center}
\end{figure}

\subsection{Transmission}

For simplicity, we discuss only zero temperature, so that
$\epsilon_k$ is set equal to the Fermi energy in the leads. The $2
\times 2$ scattering matrix is easily related to the matrix of
retarded single-particle Green functions,
$G^{\sigma}_{kk'}(\omega)$, evaluated on the energy shell,
$\omega=\epsilon_k$.\cite{hewson} We use the equation-of-motion
method to express $G^{\sigma}_{kk'}(\omega)$ and
$G^{\sigma}_{kd}(\omega)$ in terms of the Green function on the
dot, $G^{\sigma}_{dd}(\omega )$ \cite{details}. The
equation-of-motion for the latter has the form
\begin{eqnarray}
(\omega -\epsilon_{d})G^{\sigma}_{dd}(\omega )=1+\sum_{k}{\cal
V}_{k}G^{\sigma}_{kd}(\omega )+U\Gamma^{\sigma}_{ddd,d}(\omega ).
\label{GG}
\end{eqnarray}
The last term on the RHS represents the effect of the
interactions, with $\Gamma^{\sigma}_{ddd,d}(\omega )$ being the
temporal Fourier transform of $\Gamma^{\sigma}_{ddd,d}(t)=-i\Theta
(t)\langle\bigl
[d^{\dagger}_{\overline{\sigma}}(t)d_{\overline{\sigma}}(t)d_{\sigma}(t),
d^{\dagger}_{\sigma}\bigr ]\rangle $. The second term on the RHS
of Eq. (\ref{GG}) can be written as $\sum_{k}{\cal
V}_{k}G^{\sigma}_{kd}(\omega )=-A(\omega )G^{\sigma}_{dd}(\omega
)$, implying that $-A(\omega)$ contains the full contribution of
the non-interacting Hamiltonian to the self-energy on the dot.
Thus we write
\begin{eqnarray}
G^{\sigma}_{dd}(\omega )^{-1}=\omega
-\epsilon_{D}-\Sigma^{\sigma}_{\rm int}(\omega )+A(\omega
),\label{gdd}
\end{eqnarray}
where $\Sigma^{\sigma}_{\rm int}(\omega )$ represents the self
energy due to the interactions, which vanishes when $U=0$. At
$\omega=\epsilon_k$ we have
\begin{eqnarray}
A(\epsilon_k)&=e^{i|k|}\frac{J_\ell^2+J_r^2}{J}\big
[1-\frac{t_Be^{i|k|}}{2 i \sin |k|} \big (1+\gamma_D \cos \phi
\big )\big ]\nonumber\\
&=\frac{J_\ell^2+J_r^2}{J}[e^{i|k|}+ \sin |k|(Z+iY)],
\end{eqnarray}
where the $\phi$-dependent quantities $Y$ and $Z$ are defined in
this equation, while $\gamma_D=2J_\ell J_r/(J_\ell^2+J_r^2)$ is
the asymmetry factor for the dot, and
\begin{eqnarray}
t_B=-i \sin \delta_B e^{i \delta_B}=2i V_B\sin |k|/(J+2 V_B
e^{i|k|})
\end{eqnarray}
is the transmission amplitude of the ``background" path (when
$J_\ell=J_r=0$), with the effective hopping energy
$V_B=I^2/(\epsilon_k-\epsilon_0)$.

The other equations of motion\cite{details} then yield the other
Green functions, and we end up with
\begin{eqnarray}
t^\sigma&=t_B G^{\sigma}_{dd}(\epsilon_k)\bigl [\frac{J_\ell
J_r}{V_B}
e^{i\phi}+G^{\sigma}_{dd}(\epsilon_k)^{-1}-A(\epsilon_k)\big
]\nonumber\\
&=t_B G^{\sigma}_{dd}(\epsilon_k)\bigl [\frac{J_\ell J_r}{V_B}
e^{i\phi}+\epsilon_k-\epsilon_D-\Sigma^{\sigma}_{\rm
int}(\epsilon_k)\big ]. \label{tt1}
\end{eqnarray}
Equation (\ref{tt1}) is one of our main results. It expresses
$t^\sigma$ in terms of the fully dressed single particle QD Green
function, which depends on both paths of the interferometer and
therefore also on the AB phase $\phi$. The remaining discussion
aims to see if one can extract information on the ``intrinsic" QD
transmission from measurements of ${\cal T}^\sigma=|t^\sigma|^2$.

The Onsager relations require that the conductance, and therefore
also ${\cal T}^\sigma=|t^\sigma|^2$, must be an even function of
$\phi$. It is clear from Eq. (\ref{tt1}) that this holds only if
\begin{eqnarray}
&\Im[G^{\sigma}_{dd}(\epsilon_k)^{-1}-A(\epsilon_k)]
\nonumber\\
&=\Im[\epsilon_k-\epsilon_D-\Sigma^{\sigma}_{\rm
int}(\epsilon_k)]\equiv 0. \label{real}
\end{eqnarray}
Indeed, we found the same condition to follow from the {\bf
unitarity} of the scattering matrix.
The same sort of
relation appears for the single impurity scattering, in connection
with the Friedel sum rule\cite{langreth}. Equation (\ref{real})
implies that the interaction self-energy $\Sigma^{\sigma}_{\rm int
}(\epsilon_k)$ is real, and is an even function of $\phi$. It also
implies that $\Im G^{\sigma}_{dd}(\epsilon_k)^{-1}$
is {\bf fully determined by the non-interacting self-energy} $\Im
A(\epsilon_k)$.

It is now convenient to rewrite $G^{\sigma}_{dd}(\epsilon_k)$ in
terms of its phase, $\delta$. Writing
$[G^{\sigma}_{dd}(\epsilon_k)\sin |k|
(J_\ell^2+J_r^2)/J]^{-1}=(1+Y)(i-\cot \delta)$, we find
\begin{eqnarray}
{\cal
T}^{\sigma}&=|t^\sigma|^2=\frac{\sin^{2}\delta}{(1+Y)^{2}}\Bigl [
\tilde{\gamma}^{2}_{D}\sin^{2}\phi +\Bigl (\tilde{\gamma}_
D\cos\phi\nonumber\\
&-\sqrt{T_{B}}\bigl ((1+Y)\cot\delta+\cot |k|+Z\bigr )\Bigr
)^{2}\Bigr ],\label{TT}
\end{eqnarray}
with $\tilde \gamma_D=\gamma_D J/|J+2V_B
e^{i|k|}|=\gamma_D|\sin(\delta_B+|k|)/\sin|k||$ and $T_B=|t_B|^2$.
Interestingly, the second term in the square brackets is of the
Fano form\cite{fano}. At $\phi=0$, it reflects the possibility for
a complete destructive interference, with ${\cal T}^\sigma=0$. A
similar expression for ${\cal T}$ was derived by Hofstetter {\it
et al} \cite{H}, but their approximations ignore the explicit
dependence of some parameters (e.g. $Y$) on $\phi$.

When one cuts off the reference path, $V_B=0$, Eq. (\ref{tt1})
reduces to the ``intrinsic" QD transmission amplitude,
$t_D^\sigma=-i \gamma_D\sin \alpha e^{i\alpha}$, with
\begin{eqnarray}
-\cot\alpha=\cot
|k|+\frac{\epsilon_k-\epsilon_D-\Sigma^\sigma_{\rm
D,int}(\epsilon_k)}{\sin|k|(J_\ell^2+J_r^2)/J}, \label{alpha}
\end{eqnarray}
where $\Sigma^\sigma_{\rm D,int}(\epsilon_k)=\Sigma^{\sigma}_{\rm
int}(\epsilon_k)|_{V_B=0}$. It is interesting to note that for our
single-channel (one-dimensional) leads, we have ${\cal
T}_D^\sigma=\gamma_D^2 \sin^2\alpha$ (as already noted by Ng and
Lee\cite{ng}). Since $\gamma_D$ does not depend on the energy
$\epsilon_k$ or on the gate voltage $V=\epsilon_D$, it follows
that {\bf a measurement of ${\cal T}_D$ immediately also yields
the phase} $\alpha$, via $\sin\alpha=\sqrt{{\cal T}_D/\max({\cal
T}_D)}$! In many cases, the measurement of ${\cal T}_D$ already
yields $\alpha$, eliminating the need to perform complicated
interferometer measurements.  It would be very interesting to test
this, for cases where $\alpha$ is measured independently (e.g.
with an open interferometer, see below).

The phase $\delta$ of $G^{\sigma}_{dd}(\epsilon_k)$ is now given
by
\begin{eqnarray}
\cot\delta=(\cot\alpha-Z+X)/(1+Y), \label{delta}
\end{eqnarray}
where $X=(\Sigma^{\sigma}_{\rm
int}(\epsilon_k)-\Sigma^{\sigma}_{\rm
D,int}(\epsilon_k))/[\sin|k|(J_\ell^2+J_r^2)/J]$ contains only the
effects of the reference path on the interaction self-energy of
the dot. One way to proceed is to calculate $X$, e.g. using a
perturbative expansion in $V_B$ or using a full solution of the
interacting case. Even without calculating $X$, we expect it to be
an even function of $\phi$. Although $\alpha$ does not depend on
$\phi$, $\delta$ usually depends on $\phi$ (via $X,~Y$ and $Z$).

Equations (\ref{TT},\ref{alpha},\ref{delta}) represent our second
main result: they relate the $\phi$-dependence of the measured
${\cal T}^\sigma$ with the QD parameters. We are not aware of
earlier discussions which separate between the roles played by
$\alpha$ and $\delta$.

\subsection{Possible measurements}

We now discuss a few limits in which a measurement of ${\cal
T}^\sigma$, Eq. (\ref{TT}), can yield information on the
``intrinsic" phase $\alpha$ and thus on the full ``intrinsic"
transmission $t_D$. First, consider a relatively {\bf open} dot,
with small barriers at its contacts with the leads. In such
circumstances,
the electron wave function spreads over the leads, and it has only
a small amplitude on the dot itself, thus reducing the effects of
interactions. This, and the related better screening\cite{mat},
imply that $|X|\ll |Z|$. In this limit, Eq. (\ref{delta}) (with
$X=0$) gives the detailed dependence of $\delta$ on $\phi$.
Equation (\ref{TT}) then has the form
\begin{eqnarray}
{\cal T} &\equiv |t|^2=A\Big |\frac{e^{i\phi}+K}{1+z\cos\phi}\Big
|\nonumber\\
&=A \frac{1+K^2+2K \cos\phi}{1+2\Re z\cos\phi+|z|^2\cos^2\phi},
\label{T}
\end{eqnarray}
with coefficients which depend on $\gamma_D,~\delta_B,~|k|$ and
$\alpha$. For the closed interferometer, $K$ is real. The
$\cos\phi$ in the denominator, which results from the self energy
in the dot Green function,  is due to interference within the
ring, between clockwise and counterclockwise motions of the
electron. In the limit of small $V_B$ one has $z \propto t_D$.
Fitting data to Eq. (\ref{T}) would confirm that one is in this
non-interacting limit, and would yield the $V$-dependence of the
``intrinsic" phase $\alpha$ (as well as the $V$-independent
parameters $\gamma_D,~\delta_B$ and $|k|$). Note that a fit to the
explicit $\phi$-dependence in Eq. (\ref{T}) is much preferred over
a fit to a harmonic expansion of the form ${\cal T}=\sum a_n \cos
n\phi$. Figure 3 shows an example of the $V$- and
$\phi$-dependence of ${\cal T}$ for this limit. The denominator in
Eq. (\ref{T}), which contains the information on $\alpha$ via $z$
(and thus determines the higher harmonics in ${\cal T}$), is
mostly visible near the resonance at $V=0$. Far away from the
resonance this denominator is small, and ${\cal T}$ can be
approximated by $\bar{A}+B\cos(\phi+\beta)$, where $\beta$ is zero
or $\pi$ below or above the point $V=0$.

\begin{figure}[h]
\begin{center}
\leftline{\epsfclipon\epsfxsize=3in\epsfysize=3in\epsffile{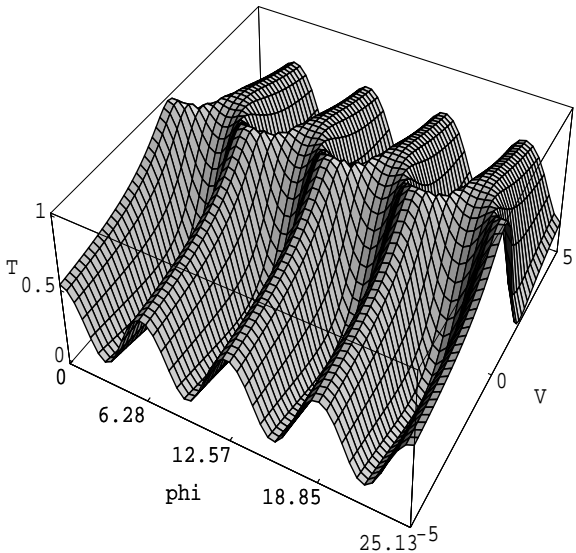}}
\caption{Dependence of the AB transmission ${\cal T}$ on the AB
phase $\phi$ and the gate voltage $V$ for a single non-interacting
resonance.}\label{fig3}
\end{center}
\end{figure}

Second, note that all the $\phi$-dependent functions $X,~Y$ and
$Z$ become small when $V_B$ (and therefore $\delta_B \sim V_B/J$)
is small. Except very close to the QD resonance, where
$\alpha=\pi/2$, we conclude that $\delta =\alpha+{\cal
O}(\delta_B)$. Thus, at least away from resonances one can study
the transmission for several values of $V_B$ (which can be varied
via the point contact voltage $\epsilon_0$), estimate $\delta(V)$
(as some average over $\phi$) for each $V_B$ and extrapolate $V_B$
to zero to obtain $\alpha$, without explicit knowledge of
$X(\phi)$.

The situation becomes even more interesting in the Kondo regime,
when $\alpha$ and/or $\delta$ remain close to $\pi/2$ over a wide
range of $V$. Hofstetter {\it et al.}\cite{H} assumed that
$\delta=\pi/2$, and deduced the $\phi$-dependence of ${\cal
T}^\sigma$. As seen from Eq. (\ref{delta}), one cannot have
$\delta=\pi/2$ without neglecting the $\phi$-dependence of $X,~Y$
and $Z$. Alternatively, if the Kondo condition implies that
$\alpha=\pi/2$ then the measurement of the $\phi$-dependence of
${\cal T}^\sigma$ could give information on the $\phi$-dependence
of $X$. If $V_B$ is sufficiently small then $\delta$ will show
small $\phi$-dependent deviations from $\pi/2$ within the Kondo
plateau, and remain very close the $\alpha$ elsewhere. For larger
$V_B$, a resonance on the QD (i.e. $\alpha=\pi/2$) may result in
significant deviation of $\delta$ from $\pi/2$. The question
whether this deviation relates to the variety of plateau values
observed by Ji {\it et al.}\cite{ji} deserves further study.

In any case, the above discussion demonstrates that although a
measurement of the transmission of the closed AB interferometer
does not yield a non-trivial AB phase shift $\beta$, the data
still contain much information on the properties of the
``intrinsic" QD.

\section{Open AB interferometer}

\subsection{Model for multi-resonances}

We next discuss the conditions for obtaining the equality
$\beta=\alpha$ in an open AB interferometer. In principle, we
could repeat the above discussion for the open case; as seen
below, this simply amounts to replacing each ``lossy" channel by a
complex self-energy at its ``base", and then proceeding as in the
closed case. However, for the present purpose it suffices to
discuss an effectively non-interacting case.\cite{bih} Obviously,
if one cannot achieve this aim in that case then there will be
problems also in the more complicated interacting cases. We thus
restrict this discussion to an approximate treatment of the
Coulomb blockade (and not the Kondo) region, and treat the
interaction term in the Hartree approximation, replacing
$n_{d\sigma}n_{d\overline{\sigma}}$ by $\langle
n_{d\overline\sigma}\rangle n_{d\sigma}$. We further simplify this
by replacing $\langle n_{d\overline\sigma}\rangle$ by a constant
which increases for consecutive resonances. Thus, we replace the
single QD by a set of smaller dots, each containing a single
resonant state, with energy $\epsilon_{D}=E_R(n)\equiv
V+U(n-1),~n=1,...,N \}$. Each such state is connected to its left
and right nearest neighbors (nn's) on the leads (denoted by $L$
and $R$) via bonds with hopping amplitudes $\{
-J_L(n),~-J_R(n),~n=1,...,N \}$ (see Fig. 4). The problem now
reduces to a simple tight-binding model, and for the ``intrinsic"
QD we find
\begin{eqnarray}
t_{D}=\frac{S_{LR} 2 i \sin k}
{(S_{LL}+e^{-ik})(S_{RR}+e^{-ik})-|S_{LR}|^2}, \label{tintr}
\end{eqnarray}
where $S_{XY}\equiv
\sum_{n}J_{X}(n)J_Y(n)^\ast/[\epsilon_k-E_R(n)]/J,~X,Y=L,R$.

\begin{figure}[h]
\begin{center}
\leftline{\epsfclipon\epsfxsize=2.5in\epsfysize=2in\epsffile{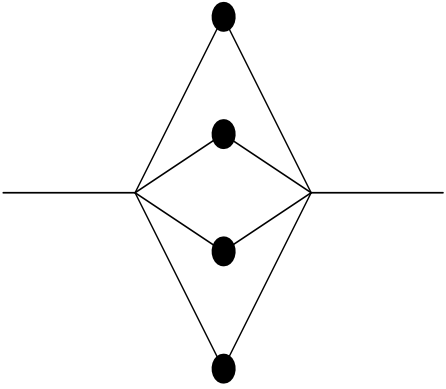}}
\vspace{.5cm}
\caption{Model for a QD with four resonances (to
replace the single dot in Fig. 2).}\label{fig4}
\end{center}
\end{figure}

Figure 5 shows typical results for the ``intrinsic" transmission
${\cal T}_{D}$ and phase $\alpha$, where the zero of $\alpha$ is
set at its ($k$-dependent) value at large negative $V$. Here and
below, we choose $k=\pi/2$, so that $\epsilon_k=0$ and the
resonances of the transmission, where ${\cal T}_{D}=1$, occur
exactly when $E_R(n)=\epsilon_k=0$, i.e. when $V=-U(n-1)$. We also
use the simple symmetric case, $J_L(n)=J_R(n) \equiv J$, and
measure all energies in units of $J$. Interestingly, this model
reproduces the behavior apparently observed by Schuster {\it et
al.}\cite{schuster}: $\alpha$ grows smoothly from 0 to $\pi$ as
$\epsilon_k$ crosses $E_R(n)$, and exhibits a sharp ``phase lapse"
from $\pi$ to 0 between neighboring resonances, at points where
${\cal T}_{D}=0$. These latter points, associated with zeroes of
$S_{LR}$, represent Fano-like destructive interference between the
states on the QD \cite{fano,ryu,jlt}. These zeroes in ${\cal T}$
are missed, and the corresponding ``phase lapses" are
superfluously smeared, when one replaces the exact Eq.
(\ref{tintr}) by a sum of Breit-Wigner resonances\cite{hack}.

\begin{figure}[h]
\begin{center}
\leftline{\epsfclipon\epsfxsize=2.5in\epsfysize=3in\epsffile{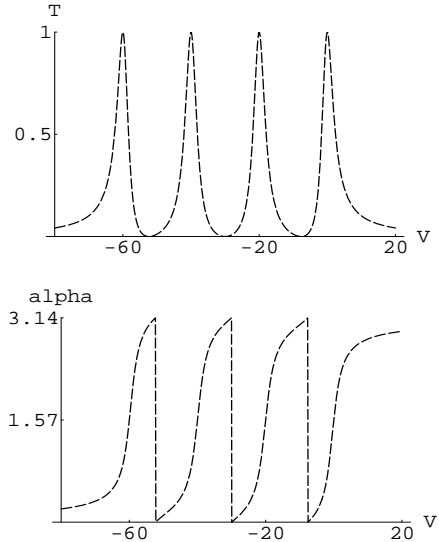}}
\caption{``Intrinsic" transmission ${\cal T}_D$ and phase
$\alpha$, for $N=4$ resonances, with $U=20J$, $J_L(n)=J_R(n)=J$
and $k=\pi/2$. The gate voltage $V$ is measured in units of $J$.
}\label{fig5}
\end{center}
\end{figure}

It is interesting to note that, although there are several
resonances, the upper part of Fig. 5 is fully reproduced when one
takes $\alpha$ from the lower part, and plots $\sin^2\alpha$. This
emphasizes again the possibility to {\bf measure} $\alpha$ {\bf
directly} from ${\cal T}$.

\subsection{Model for the open AB interferometer}

Placing the above QD model on one branch of the closed AB
interferometer, one can easily solve for ${\cal T}$. In the
absence of interactions, we again find the form (\ref{T}), which
is thus also valid for many resonances. Near each resonance,
results are similar to those shown in Fig. 3. As expected, Eq.
(\ref{T}) is still an even function of $\phi$, with no phase shift
$\beta$ (except for the apparent jumps between zero and $\pi$).

Before we discuss the open interferometer, it is useful to
understand the criteria for having the two-slit situation. A
crucial condition for having $t=t_De^{i\phi}+t_B$ is that the
electron go through each branch only {\bf once}. Equivalently,
there should be {\bf no reflections} from the ``forks", which
connect the ring with the external leads, back into the
ring\cite{prl}. We achieve this by the construction shown in Fig.
1c: each of the four ``lossy" channels in Fig. 1b is now replaced
by a ``comb" of $M$ channels. An analysis of each such ``comb"
shows that (a) the transmission $T$ and reflection $R$ through the
``comb" is only weakly dependent on the energy $\epsilon_k$ near
the band center $k=\pi/2$ (Fig. 6a) and (b) the transmission $T$
decreases and the reflection $R$ increases with the coupling of
each ``lossy" channel in the ``comb" to its ``base", $J_x$ (Fig.
6b) and with the number $M$.\cite{bih} Note that $T+R<1$, due to
the loss into the ``teeth" of the ``comb". Ideally, one would like
to have $T,~R\ll 1$, so that in practice the electron crosses each
``comb" only once, and is not reflected from the ``comb" back into
the QD. Given Fig. 6b, we expect to have optimal two-slit
conditions for intermediate values of $J_x$, e.g. $J_x \sim 0.9J$
for $M=6$.

\begin{figure}[h]
\begin{center}
\begin{tabular}{cc}
{\epsfxsize=1.5in\epsfbox{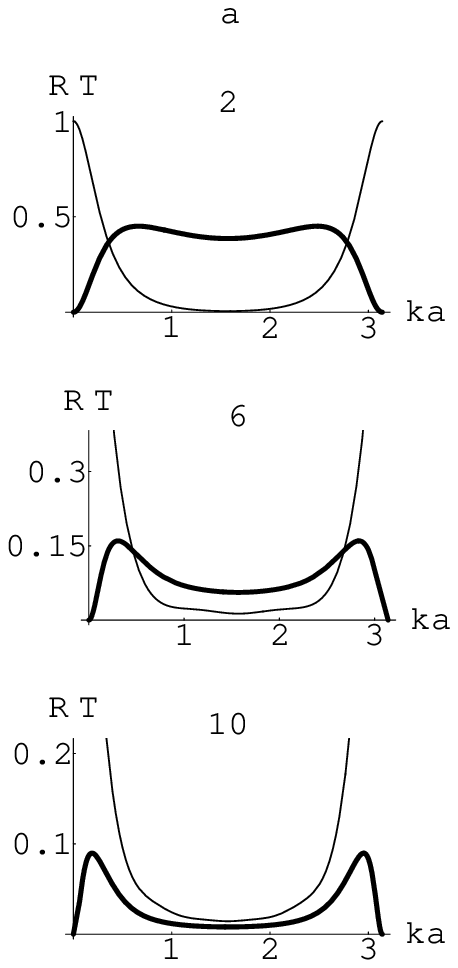}}&
{\epsfxsize=1.5in\epsfbox{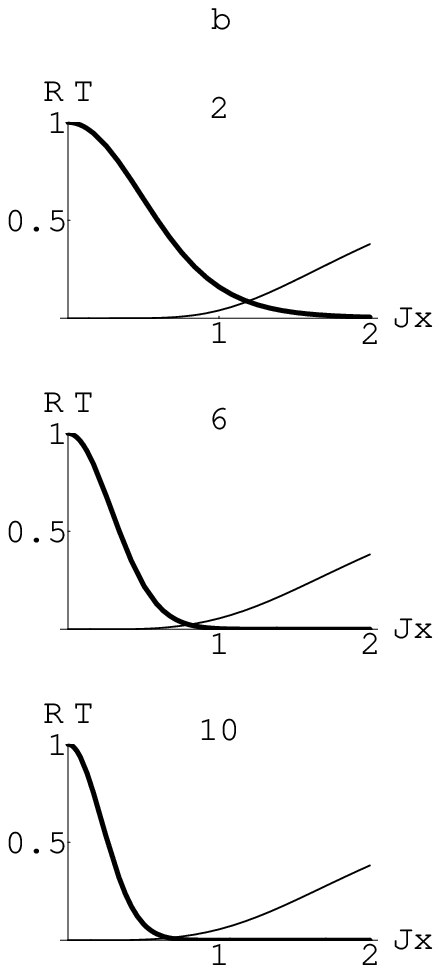}}
\end{tabular}
\vspace{0.3cm} \caption{Transmission (thick line) and reflection
(thin line) through a ``comb", (a) versus $k$ at $J_x=.7J$ and (b)
versus $J_x$ (in units of $J$) at $k=\pi/2$. The number on each
frame gives the number of ``teeth", $M$.}\label{fig6}
\end{center}
\end{figure}

We next present results for the open interferometer shown in Fig.
1c. The algebra is similar to that for the closed case: each
``tooth" of the ``combs" can be eliminated from the equations, at
the cost of adding a self energy on the site of its ``base", equal
to $J_x^2e^{ik}/J$. The result is again of the form of the first
expression in Eq. (\ref{T}), except that now $K$ becomes complex
and therefore the numerator in the second expression contains
$\cos(\phi+\gamma)$ instead of $\cos\phi$.  To demonstrate
qualitative results, we chose four identical ``combs", with $M=6$
``teeth" on each and with the same hopping matrix elements; all
the hopping energies were set to $-J$, except for the bonds
connecting each ``tooth" to its base, denoted $-J_x$. Figure 7
shows exact results for $A,~B,~C$ and $\beta$ in fits of ${\cal
T}$ to the form
\begin{eqnarray}
T=A+B\cos(\phi+\beta)+C\cos(2 \phi+\gamma)+\ldots, \label{fit}
\end{eqnarray}

\begin{figure}[h]
\begin{center}
\begin{tabular}{cc}
\epsfbox{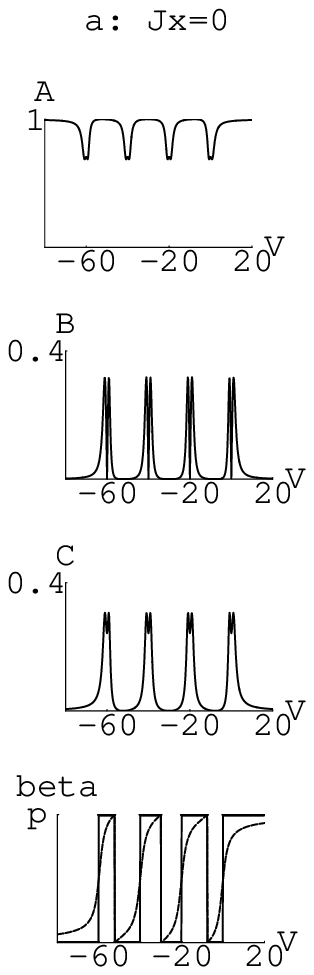}& \epsfbox{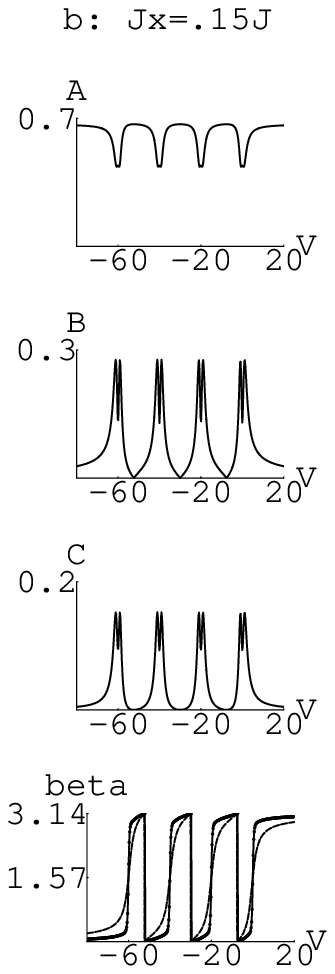}\\
\epsfbox{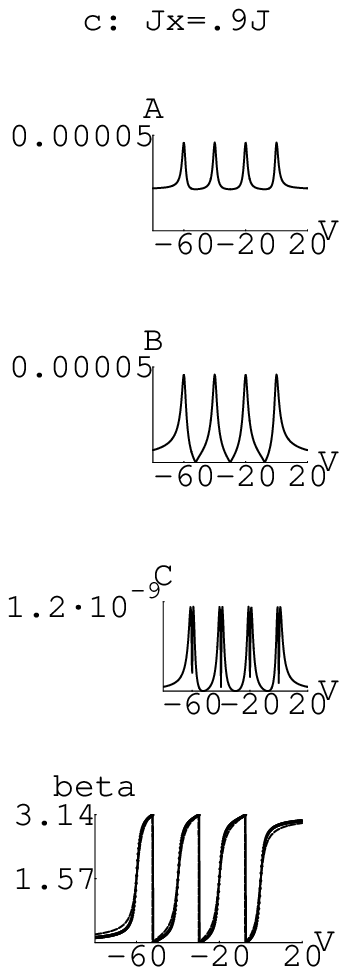}&\epsfbox{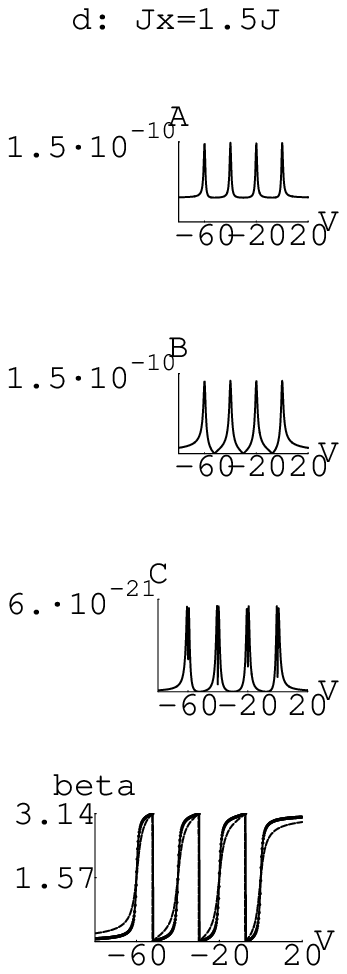}

\end{tabular}
\vspace{.74cm}

\caption{$A,~B,~C$ and $\beta$ for transmission through (a) the
closed AB ring ($J_x=0$), and for the open interferometer with (b)
$J_x=.15J$, (c) $J_x=.9J$ and (d) $1.5J$. The thin line in the
lowest frames shows the exact intrinsic phase $\alpha$ (from Fig.
5). The QD parameters are the same as in Fig. 5. }\label{fig7}
\end{center}
\end{figure}

\noindent with the conventions $B,~C>0$, as used in the analysis
of experiments. Note the decreasing magnitude of the amplitudes as
$J_x$ increases, due to the large losses to the lossy channels.
Note also that for $J_x=0$ the results reproduce those for the
closed interferometer, with $\beta$ jumping discontinuously
between 0 and $\pi$. As $J_x$ increases, the ``data" for $\beta$
become smoother, and they approach the ``intrinsic" values of
$\alpha$ for intermediate values of $J_x \sim 0.9J$. However, as
$J_x$ increases further, $\beta$ ``retracts" towards a more steep
variation near each resonance. Although the electron crosses each
``comb" only once, due to the small values of the comb
transmission $T$, it is reflected several times from the ``combs"
back towards the QD, due to the increasing ``comb" reflection.
Therefore, the electron visits the QD several times, and the final
AB transmission does not reflect the correct desired $t_D$.

Finally, we allow also a lossy channel connected directly to the
dot. As seen in Fig. 8, this eliminates the Fano zeroes of $B$ and
causes a ``smearing" of the sharp Fano ``phase lapses" in $\beta$.
Technically, the losses from the QD introduce complex
self-energies on the dot, which move the zeroes of ${\cal T}$ away
from the real energy axis. Interestingly, Fig. 8 resembles the
experiments of Schuster {\it et al.}\cite{schuster}

\begin{figure}[h]
\begin{center}
\leftline{\epsffile{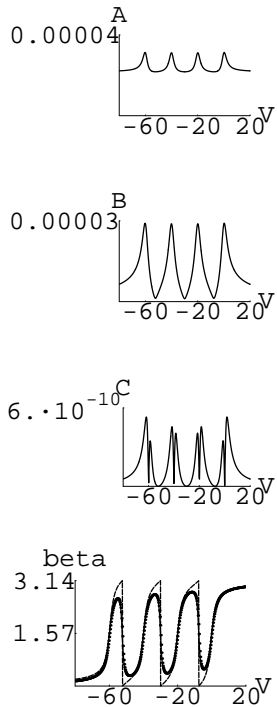}} \caption{Same as Fig. 7, but with a
``lossy" channel attached to the QD, with coupling $J_x'=J_x=.9J$.
}\label{fig8}
\end{center}
\end{figure}

\section{Conclusions}

Basically, we have made three explicit predictions:

$\bullet$ For single channel leads, the QD transmission and its
phase are related via ${\cal T}_D=\gamma^2_B\sin^2\alpha$. In such
cases, the measurement of ${\cal T}_D$ also yields $\alpha$.

$\bullet$ Measurements of the transmission ${\cal T}$ in a closed
interferometer contain much information on {\bf both} the
magnitude ${\cal T}_D$ and the phase $\alpha$ of the ``bare" QD.

$\bullet$ Open interferometers do not usually obey the two-slit
criteria. Therefore, the phase $\beta$ measured via a fit to Eq.
(\ref{fit}) will usually not yield the intrinsic QD phase
$\alpha$. However, optimization of the losses
can achieve the two-slit conditions, and yield
$\beta=\alpha$.

In principle, the configuration of Fig. 1c allows a full study of
all the cases discussed here: setting the voltage on the reference
site $\epsilon_0 \rightarrow \pm \infty$ sends $V_B$ to zero, and
yields the ``intrinsic" transmission through the QD, ${\cal T}_D$.
Setting the gate voltage on the QD $V=\epsilon_D \rightarrow \pm
\infty$ yields the reference transmission $T_B=\sin^2\delta_B$.
Setting the coupling to the ``lossy" channels ($J_x$) to zero, by
some manipulations of the relevant gates, yields the ``closed"
case. In this case, variation of $\epsilon_0$ allows variation of
$V_B$, and extrapolation of $\delta$ to $\alpha$. Finally, varying
$J_x$ allows optimization of the two-slit condition, yielding
another measurement of the ``intrinsic" phase $\alpha$. We hope
that this review will stimulate the buildup of such flexible
experimental systems.

\section*{Acknowledgements}

We thank M. Heiblum, Y. Levinson, A. Schiller, H. A.
Weidenm\"uller and A. Yacoby for helpful conversations. This
project was carried out in a center of excellence supported by the
Israel Science Foundation, with additional support from the Albert
Einstein Minerva Center for Theoretical Physics at the Weizmann
Institute of Science, and from the German Federal Ministry of
Education and Research (BMBF) within the Framework of the
German-Israeli Project Cooperation (DIP).

\end{multicols}
\end{document}